\documentclass[12pt]{article}
\usepackage[dvips]{graphicx}
\usepackage{amssymb}
\usepackage{amsmath}

\def\beq{\begin{equation}}\def\eeq{\end{equation}}
\def\bea{\begin{eqnarray}}\def\eea{\end{eqnarray}}

\begin{document}
 
\title{Realistic collapse model of bosonic strings}
 
\author{Roman Sverdlov
Institute of Mathematical Sciences ,
\\IV Cross Road, CIT Campus, Taramani, Chennai, 600 113, Tamil Nadu, India} 
\date{June 7, 2012}
\maketitle
 
\begin{abstract} In this paper we will utilize the non-trivial shapes of the strings in order to come up with realistic definition of probability amplitudes in a lot more natural way than could be done in point particle counterpart. We then go on to ''translate'' GRW model to string theory context. In this paper we limit ourselves to boson-only toy model without D-branes.
\noindent  
\end{abstract}

\subsection*{1. Introduction}

One of the key problems in interpretation of quantum mechanics can be traced to the fact that co-existence of different paths in ''path integral'' can not be accommodated ''classically''.  In case of string theory, one can attempt to address this question by utilizing already-existing branchings of strings. Conventionally, it is assumed that string branching represents "classical" emissions and re-absorption within one single trajectory, which is to be "quantized" with path integration ''later on''. In this paper we propose a different point of view. We claim that \emph{some} of the branchings represent emissions/absorptions within single trajectory while \emph{others} represent interference between different trajectories leading to path integral. 

In order to distinguish the two types of branchings, we need some further construction. We propose a "preferred" $(\sigma, \tau)$ parametrization that is globally defined across all of the string branches. We then ''slice'' the string by ''lines'' $\tau=0$, $\tau=1$, $\tau=2$, etc. The slice between $\tau=k-1$ and $\tau=k$ will consist of $a_k$ domains, $D_{k1}, \cdots, D_{ka_k}$, that are connected within themselves but disconnected from each other. The domain $D_{kl}$, itself, might consist of several branches. We claim that the branches within the domain $D_{kl}$ represent emission/absorption within a single trajectory, whereas $D_{kl}$ and $D_{kl'}$ represent two different trajectories. Despite the fact that $D_{kl}$ and $D_{kl'}$ are disconnected, they can both be connected to $D_{k+1, j}$. This connection represent the "merging" of separate trajectories at ''common endpoint''.

It should be admitted that this proposal only leads to discretized version of the path integral. After all, each "allowed" trajectory can be identified with sequence $\{b_k \}$ ($1 \leq b_k \leq a_k$) corresponding to $D_{1b_1} \cup D_{2b_2} \cup \cdots$. In order for this to approximate path integral, we need $a_k$ to be large enough so that any trajectory imaginable will be approximated by at least one of the above sets. In other words, our version of the string will be ''branching'' a lot more than it is usually assumed. This, however is only a difference in quantity rather than the quality. The only "qualitative" difference involves ''artificial'' $\tau=k$ lines. Their presence can be "justified" by "further utilizing" them for the "string version" of GRW spontaneous collapse model, which is the bulk of this paper. 

\subsection*{2. Realistic definition of probability amplitude}

Let us denote by $P_{kl}$ the set of all possible $i$ for which the domain $D_{k-1, i}$ is connected to $D_{kl}$, and let us denote by $F_{kl}$ the set of all possible $j$ for which the domain $D_{kl}$ is connected to $D_{k+1, j}$ (here ''P'' stands for ''past'' and ''F'' stands for ''future''). We then assume that wave function $\psi (\sigma, \tau)$ on the string satisfies an appropriate ''local'' dynamics (such as, for example, the one proposed in Section 4) that would lead to the emergent "global" feature
\beq \forall (\sigma, \tau) \in D_{kl} \bigg( \psi (\sigma, \tau) \approx e^{iS (D_{k,l})} \sum_{i \in P_{kl}} \frac{\int_{D_{k-1, i}} d \tau' d \sigma' \psi (\sigma', \tau')}{A (D_{k-1, i})}  \bigg) \label{NonlocalRealism} \eeq
where
\beq S (D) = \int_D d \sigma d \tau {\cal L} (X^{\mu}; \sigma, \tau) \; , \; A (D) = \int_D d \sigma d \tau \label{Integrals}\eeq
and it is assumed that $\cal L$ includes some $\delta$-functions at the boundary which would produce Euler-related terms. If we define $\psi (D)$ to be 
\beq \psi (D) = \frac{1}{A (D)} \int_{D_{k-1, i}} d \tau' d \sigma' \psi (\sigma', \tau') \label{AveragePsi} \eeq
the Equation \ref{NonlocalRealism} implies 
\beq \forall (\sigma, \tau) \in D_{kl} \forall k \geq 2 \Big(\psi (\sigma, \tau) \approx e^{iS (D_{k,l})} \sum_{i \in P_{kl}} \psi (D_{k-1},i) \Big) \eeq
Let us denote by $T_{k_1,l_1;k_2,l_2}$ (letter $T$ stand for "trajectory") all possible sequences $\{b_{k_1+1}, \cdots, b_{k_2-1} \}$ such that $D_{k_1,l_1}$ is connected to $D_{k_1+1, b_{k_1+1}}$, $D_{k_1+1, b_{k_1+1}}$ is connected to $D_{k_1+2, b_{k_1+2}}$, and so forth, $D_{k_2-2, b_{k_2-2}}$ is connected to $D_{k_2-1, b_{k_2-1}}$ and finally $D_{k_2-1, b_{k_2-1}}$ is connected to $D_{k_2,l_2}$. Furthermore, let $I_{k_1; k_2,l_2}$ (letter $I$ stands for "initial") be the set of choices of $l_1$ for which at least one such sequence exists. In this case it is easy to show by induction that 
\beq \psi (D_{k_2, l_2}) \approx \sum_{l_1 \in I_{k_1; k_2,l_2}} \bigg(\psi (D_{k_1, l_1}) \times \label{NonlocalRealism2}\eeq
\beq \times \sum_{ \{b_{k_1 +1}, \cdots, b_{k_2-1} \} \in T_{k_1, l_1; k_2, l_2}} e^{ iS (D_{k_1+1, b_{k_1+1}}) + \cdots + iS (D_{k_2-1, b_{k_2-1}}) + iS (D_{k_2, l_2}) } \bigg) \nonumber \eeq
If we were to assume that $\psi (D_{k_1, l_1}) = \delta_{l_1}^1$, then the above expression becomes identical to discretized path integral of "starting at" $D_{k_1 1}$ and "ending at" $D_{k_2 l_2}$. The value of $\psi (D_{k_2, l_2})$ represents the "probability amplitude" that the "trajectory" represented by $D_{k_2 l_2}$ (including all of the emission/absorption events represented by handles of $D_{k_2 l_2}$) "takes place". At the same time, the said probability amplitude is defined in "realistic" way; after all, equations \ref{NonlocalRealism} and \ref{AveragePsi} imply that $\psi (\sigma, \tau) = \psi (D_{k_2, l_2})$ for any $(\sigma, \tau) \in D_{k_2, l_2}$. 

\subsection*{3. String version of GRW collapse model} 

Let us now proceed to ''converting'' the GRW model of quantum measurement into string theory context. In point particle case (for more details, see \cite{GRW}) the GRW model involves multiplication of wave function (which lives in $N$-particle configuration space) by random Gaussians at randomly chosen moments in time, 
\beq \psi (\vec{x}, t^+) = K (\vec{y}, t) e^{- \frac{\alpha}{2} \vert \vec{x} - \vec{y} \vert^2} \psi (\vec{x}, t^-) \; , \; K (\vec{y}, t)= \bigg( \int d^{3N} x \; e^{- \alpha \vert \vec{x} - \vec{y} \vert^2} \vert \psi (\vec{x}, t^-) \vert^2 \bigg)^{-1} \label{NormK}\eeq
Furthermore, the probability density that $\vec{y} = \vec{y}_0$ is given by $K^{-2} (\vec{y}_0)$. It has been shown (see \cite{GRW}) that the above model indeed produces collapse of wave function as well as Born's rule. For the reasons to be seen shortly, let us alter the ''point particle'' version of GRW model before we proceed to strings. We will define a function
\beq M (a_1, \cdots, a_n; b_1, \cdots, b_n) = \max \{k \vert \forall l \in \{1, \cdots, n \} a_l b_l \leq a_k b_k \} \eeq
We will now postulate that at any time $t=k \in \mathbb{N}$, we have a random scatter of points $\{ \vec{x}_{k1}, \cdots, \vec{x}_{ka_k} \}$ and each point $\vec{x}_{kl}$ ($l \in \{1, \cdots, a_k \}$) is assigned a random parameter $\mu_{kl} \in (0, 1)$ that is chosen according to the ''biased coin'' with Gaussian probability distribution around the origin. Finally, we will replace ''spontaneous collapse'' with 
\beq \psi (\vec{x}; t=k^+) =  K (\vec{x}_{k; M ((erfc \; \mu_{k1})^{-1}, \cdots, erfc \; \mu_{ka_k}; K^{-2} (\vec{x}_{k1}), \cdots, K^{-2} (\vec{x}_{ka_k}) )}, t) \times \eeq
\beq \times  \exp (-\vert \vec{x}-  \vec{x}_{k; M ((erfc \; \mu_{k1})^{-1}, \cdots, (erfc \; \mu_{ka_k})^{-1}; K^{-2} (\vec{x}_{k1}), \cdots, K^{-2} (\vec{x}_{ka_k}) )}\vert^2)  \psi (\vec{x}; t= k^-) \nonumber \eeq
where $erf$ and $erfc$ stands for ''error function'' and ''complimentary error function'' defined by
\beq erfc \; (a) = 1 - erf \; (a) = 1- \frac{1}{\sqrt{\pi}} \int_{-a}^a e^{-x^2} dx \eeq
It is easy to see that if the values of $\mu_{kl} \in (0,1)$ are unknown then, the probability of a given point being a ''center'' of a Gaussian is proportional to $K^{-2}$ as claimed by GRW theory. 

Let us now turn to string case. The ''point'' $\vec{x}_{kl}$ will now be replaced by a domain $D_{kl}$.We will further replace $\mu_{kl}$ with a function $\mu (\sigma, \tau) \in (0, 1)$ subject to the dynamics (to be described in the next section) which would guarantee that 
\beq \forall k, l \; \forall (\sigma, \tau) \in D_{kl} \; \forall (\sigma', \tau') \in D_{kl}\;  (\mu (\sigma, \tau) = \mu (\sigma', \tau')) \eeq
Thus, $\mu_{kl}$ is a ''common value'' of $\mu (\sigma, \tau)$ assigned to every single point $(\sigma, \tau) \in D_{kl}$ which can  be formally defined as 
\beq \mu_{kl} = \frac{1}{A_{kl}} \int_{D_{kl}} d \sigma d \tau \mu (\sigma, \tau) \; , \; A_{kl} = \int_{D_{kl}} d \sigma d \tau \eeq
 We now have to define $d (D_{kl}, D_{kj})$ in order to replace $\vert \vec{x} - \vec{x}_{kl} \vert$. The ''back'' (that is, $t=k^+ -1$) boundary of $D_{kl}$ consists of certain number of ''lines'' (open strings) and ''loops'' (closed strings). In order to measure the ''distances'' in a way that can be ''mechanicized'' per construction used in next section, we would instead want to resort to open strings being used alone. In order to do that, we will  identify ''preferred points'' on each ''loop'' and ''break it'' into two ''lines'', thus treating its two parts as two separate ''open strings''.  This implies two troublesome things. First of all, we predict that the closed string is ''almost the same'' as two almost-touching open strings, despite different Euler characteristics. Secondly, we predict that two identical copies of closed string with the ''preferred points'' selected in different locations are ''far away'' from each other. 

 The answer to the first question is that preferred points are only utilized for GRW model (Sections 3 and 4) and not for measurement-free Lagrangian (Section 2). As far as measurement-free Lagrangian is concerned, we implicitly assumed that $\cal L$ includes terms with boundary-centered $\delta$-functions that would lead to path integrals that produce Euler characteristics. The only part where we stray is that we claim that measurement does not respect Euler characteristics in the same way as measurement-free evolution does. The situation of having to ''measure'' only certain degrees of freedom over others, thus violating symmetries, is quite common in measurement theories. As far as the second issue is concerned, it is true that our measurement would treat identical copies of closed strings with differently chosen ''preferred points'' as ''far away''. However, what this means is simply that the GRW process we are describing will lead to ''collapse'' in our choice of ''preferred points'' \emph{in addition to} the ''collapses'' of other parameters. In light of the fact that the location of ''preferred points'' does not figure in the Lagrangian implies that the said ''collapse'' is ''orthogonal'' to everything else we are describing and, therefore, will not lead to any detectable results. In other words, the two ''preferred points'' and their supposed ''collapse'' both constitute a ''nuisance information''. For now, we will be content with this, even though for future it might be interesting to search for theories that can avoid that. 

 We will denote the number of above-described ''open strings'' at the end of $D_{kl}$ by $N_{kl}$ and denote these strings by $E_{kl;1}, \cdots, E_{kl;N_{kl}}$. We will then define the distance as
\beq N_{ij} \geq N_{kl} \Rightarrow d^2 (D_{ij}, D_{kl})= W (N_{ij} - N_{kl}) + \min_{ a_p \neq a_q \; , \; \{a_1, \cdots, a_{N_{kl}} \} \subset \{1, \cdots, N_{ij} \} } \sum_{p=1}^{N_{kl}} d^2 (E_{ijp}, E_{klp}) \eeq
\beq N_{ij}<N_{kl} \Rightarrow d^2 (D_{ij}, D_{kl})= W (N_{kl} - N_{ij}) + \min_{ a_p \neq a_q \; , \; \{a_1, \cdots, a_{N_{ij}} \} \subset \{1, \cdots, N_{kl} \} } \sum_{p=1}^{N_{ij}} d^2 (E_{ijp}, E_{klp}) \eeq
where, in $(+, -,  \cdots, -)$ convention, 
\beq d(E_{ijp}, E_{klq}) = - X^{\mu} (E_{ijp}, E_{klq}) X_{\mu} (E_{ijp}, E_{klq}) \; , \;  X^{\mu} (E_{ijp}, E_{klq}) = \int d \rho (E^{\mu}_{ijp}-  E_{klq}^{\mu}) \eeq
Thus, we would like $W$ to be sufficiently large so that creating new string is ''more difficult'' than displacing it by ordinary distance. Nevertheless, since $W$ is still finite, displacing strings on ''extremely large'' distances happens to somehow be ''more difficult'' than merely creating them. This might seem a bit paradoxical since we can always achieve displacement of the string by a large distance through its annihilation at one point and the creation at the other point. Nevertheless, despite the apparent ''poor motivation'' our definition of ''distance'' is mathematically consistent. One should keep in mind that the only place where ''distance'' is used is the GRW model. In light of the fact that the collapse model is poorly studied anyway, it is more or less okay to define the distance in the way that we did since it can't be easily falsified. Nevertheless, in future, we might explore other definitions of distance. 

We will now define the ''string version'' of Equation \ref{NormK} to be
\beq K_{k,l} = \bigg( \sum_j e^{- \alpha d^2 (D_{kl}, D_{kj})} \bigg\vert \frac{1}{A_{k-1,j}}\int_{D_{kj}} d \sigma d \tau \psi (\sigma, \tau) \bigg\vert^2 \bigg)^{-1} \eeq
Finally, we define the GRW modification of Equation \ref{NonlocalRealism} to be 
\beq \forall (\sigma, \tau) \in D_{kl} \bigg( \psi (\sigma, \tau) \approx  K_{k; M ((erfc \; \mu_{k1})^{-1}, \cdots, (erfc \; \mu_{ka_k})^{-1}; K^{-2}_{k1}, \cdots, K^{-2}_{ka_k} )} \times \eeq
\beq \times  \exp (- d^2(D_{kl}, D_{k; M ((erfc \; \mu_{k1})^{-1}, \cdots, (erfc \; \mu_{ka_k})^{-1}; K^{-2} (\vec{x}_{k1}), \cdots, K^{-2} (\vec{x}_{ka_k}) )} ) e^{iS (D_{k,l})} \sum_{i \in P_{kl}} \frac{\int_{D_{k-1, i}} d \tau' d \sigma' \psi (\sigma', \tau')}{A (D_{k-1, i})}  \bigg) \nonumber \eeq

\subsection*{4. Local dynamics that ''generates'' nonlocal results}

Let us now propose a local mechanism of generating non-local laws described above. As we mentioned earlier, on every closed string we single out two ''preferred'' points, thus turning it into open string. We will now connect any given ''open string'' in the above sense by a ''bridge'' to every single other ''open string''. Here, by a ''bridge'' we mean a surface
\beq B^{\mu}_{k^+-1;nl;mj} (\rho, \eta) = \eta E^{\mu}_{k^+-1;nl} (\rho) + (1-\eta) E^{\mu}_{k^+-1; mj} (\rho) \eeq
Where $E^{\mu}_{k^+-1;nl}$ and $E^{\mu}_{k^+-1;mj}$ describe $n$-th and $m$-th open strings on ''past'' edges of $D_{kl}$ and $D_{kj}$, respectively and $\rho \in [0, 2 \pi]$ parametrizes each edge \emph{separately} as contrasted with $\sigma \in [0, 2 \pi]$ which parametrizes entire $\tau = const$ slice (in other words two different points on  $\tau = const$ slice can \emph{not} have the same $\sigma$ but they \emph{can} have the same $\rho$ provided they are on two different ''edges''; this leads to some discontinuities in $\sigma$, but in light of $\sigma$-covariance they seem inconsequential). We will now introduce a new ''time'' coordinate $\xi$ while treating previously existing ''time'' coordinate $\tau$ as ''space''. We will postulate ''diffusion processes'' that involves motion in both $+ \tau$ and $- \tau$ directions in ''time'' $\xi$, and leads to desired outcome at the $\xi \rightarrow \infty$ equilibrium.
\[ \frac{d S}{d \xi}= \left\{ \begin{array}{ll}
         \partial_{\sigma}^2 S + \partial_{\tau}^2 S + \epsilon {\cal L} & \mbox{if $(\sigma, \tau) \in D_{kl} \setminus \partial D_{kl}$};\\
        n^{\tau} \partial_{\tau} S  + n^{\sigma} \partial_{\sigma} S- \epsilon S & \mbox{if $(\sigma, \tau) \in \partial D_{kl}$}.\end{array} \right. \]
\[ \frac{d \mu}{d \xi}= \left\{ \begin{array}{ll}
         \partial_{\sigma}^2 \mu + \partial_{\tau}^2 \mu  & \mbox{if $(\sigma, \tau) \in D_{kl} \setminus \partial D_{kl}$};\\
        n^{\tau} \partial_{\tau} \mu  + n^{\sigma} \partial_{\sigma} \mu & \mbox{if $(\sigma, \tau) \in \partial D_{kl}$}.\end{array} \right. \]
\[ \frac{d \nu}{d \xi}= \left\{ \begin{array}{ll}
         \partial_{\eta}^2 \nu + \partial_{\tau}^2 \nu  & \mbox{if $(\rho, \eta) \in B^{\mu}_{k^+-1;la;jb} \setminus \partial B^{\mu}_{k^+-1;la;jb}$};\\
        n^{\tau} \partial_{\tau} \nu  + n^{\eta} \partial_{\eta} \nu - \epsilon \nu + \epsilon \mu (\sigma (\rho, \eta), k^+-1) & \mbox{if $(\rho, \eta) \in \partial B^{\mu}_{k^+-1;la;jb} $}.\end{array} \right. \]
\[ \frac{d \chi}{d \xi}= \left\{ \begin{array}{ll}
         \partial_{\eta}^2 \chi + \partial_{\tau}^2 \chi  & \mbox{if $(\rho, \eta) \in B^{\mu}_{k^+-1;la;jb} \setminus \partial B^{\mu}_{k^+-1;la;jb}$};\\
        n^{\tau} \partial_{\tau} \chi  + n^{\eta} \partial_{\eta} \chi - \epsilon \chi + \epsilon \psi (\sigma (\rho, \eta), k^+-1) & \mbox{if $(\rho, \eta) \in \partial B^{\mu}_{k^+-1;la;jb} $}.\end{array} \right. \]
At the same time, we are assuming that the structure of string itself does not evolve in $\xi$. Thus, 
\beq X^{\mu} (\xi,\sigma, \tau) = X^{\mu} (\sigma, \tau) \; , \; B^{\mu} (\xi, \rho, \eta) = B^{\mu} (\rho, \eta) \eeq
These processes will allow a given point to ''access'' the ''global'' information used in previous section ''locally'', as long as we assume that $\xi$ is ''large enough'' for $\xi \rightarrow \infty$ equilibrium to have been reached up to very good approximation. However, there is still ''infinitesimal nonlocality'' in that you might be infinitesimally displaced on one direction from the boundary line while trying to access information infinitesimally in another direction; due to the fact that things can't diffuse across the line there is discontinuity. Nevertheless we can claim to have avoided ''finite'' nonlocality. The ''prescription'' of ''making dynamics local'' involves replacing the ingredients in the equations found in the previous section with ''local'' quantities per the following prescription:
\beq d^2 (E_{k^+-1;ip}, E_{k^+-1;jq}) \longrightarrow - (\partial_{k^+-1;ln;ip} B^{\mu} - \partial_{k^+-1;ln;jq} B^{\mu})(\partial_{k^+-1;ln;ip} B_{\mu} - \partial_{k^+-1;ln;jq} B_{\mu})\eeq
\beq \min_{ a_p \neq a_q \; , \; \{a_1, \cdots, a_{N_{k^+-1,l}} \}} \longrightarrow \min_{(i,p) \neq (j,q) \; , \; \nu_{(\sigma,k-1)+ \delta_{k^+-1;ln,ip}} = \nu_{(\sigma,k-1)+ \delta_{k^+-1;ln,jq}}}\eeq
\beq \frac{1}{A_{kj}} \int_{D_{kj}} d \sigma d \tau \psi (\sigma, \tau) \longrightarrow \psi ((\sigma, k^+-1) + \delta_{k^+-1;ln,jq})  \eeq
\beq \frac{1}{A_{k-1,i}} \int_{D_{k-1,i}} d \sigma d \tau \psi (\sigma, \tau)  \longrightarrow \psi (\sigma, k^+-1)  \eeq
\beq S(D_{kl}) \longrightarrow S(\sigma, k^+-1) \; , \; \mu_{kj} \longrightarrow 2 \nu ((\sigma, k^+-1) + \delta_{k^+-1;ln,jq}) - \mu (\sigma, k^+-1) \eeq
\[ \psi = (\cdots) \longrightarrow \frac{d \psi}{d \xi}= \left\{ \begin{array}{lll}
         \partial_{\sigma}^2 \psi + \partial_{\tau}^2 \psi & \mbox{if $(\sigma, \tau) \in D_{kl} \setminus \partial D_{kl}$};\\
        n^{\tau} \partial_{\tau} \psi  + n^{\sigma} \partial_{\sigma} \psi + \epsilon \times (\cdots)  & \mbox{if $\tau = k^+-1$} \\
        n^{\tau} \partial_{\tau} \psi  + n^{\sigma} \partial_{\sigma} \psi - \epsilon \psi  (\sigma, \tau) & \mbox{if $(\sigma, \tau) \in \partial D_{kl}$ and $k^+ -1< \tau  \leq k^-$}.\end{array} \right.  \]
Here, $\partial_{k;ln;ip}$ is the derivative in the direction ''into'' the bridge $B_{k;ln;ip}$taken at the boundary of the latter, and $\delta_{k^+-1;ln;ip}$ is a displacement in that same direction. After having made the above changes we realize that a given point ''locally'' does not ''know'' that a given direction (which we just called $ln;ip$) leads ''from'' $n$-th edge of $D_l$ "to" $p$-th edge of $D_i$. After all, this is a "global" information that is not accessible to a given point. The only thing that the point "knows" is that there is certain number of "bridges" on which edges it sits which lead to "some" direction. Thus, one more substitution is in order:
\beq \partial_{ln; ip} \longrightarrow \partial_i \; , \; \delta_{ln; ip} \longrightarrow \delta_i \eeq
It should be noted that the above might have unwanted extra coefficients proportional to the circumferences of the boundaries. These coefficients, however, will statistically be expected to approximate constants due to the ''very large'' number of strings on each ''slice''. Once we apply the above substitutions to the dynamics in Section 3, we produce
\beq N (\sigma, k^+-1; \mu) = \sharp \{j\vert \vert 2 \nu ((\sigma, k^+-1) + \delta_{k^+-1;j}) - \mu (\sigma, k^+-1) - \mu \vert < \epsilon \} \eeq
\beq  N (\sigma, k^+-1; \mu' ) \geq N (\sigma, k^+-1; \mu (\sigma, k^+-1))\Longrightarrow \nonumber \eeq
\beq \Longrightarrow d^2 (\sigma,  k^+-1; \mu')= W (  N (\sigma,  k^+-1; \mu' ) - N (\sigma,  k^+-1; \mu (\sigma,  k^+-1)) + \nonumber \eeq
\beq + \min_{ \{j_1, \cdots, j_{N (\sigma, k^+-1, \mu)} \vert \forall i \leq N (\sigma,  k^+-1, \mu (\sigma,  k^+-1)) \; , \; \vert 2 \nu ((\sigma,  k^+-1)+ \delta_{ k^+-1;j}) - \mu (\sigma,  k^+-1) -  \mu'  \vert < \epsilon \}  } \eeq
\beq \sum_{j=1}^{N (\sigma, k^+-1, \mu')} (\partial_{ k^+-1;j} B^{\mu} - \partial_{ k^+-1;j} B^{\mu})(\partial_{ k^+-1;j} B_{\mu} - \partial_{ k^+-1;j} B_{\mu})\nonumber \eeq
\beq  N (\sigma,  k^+-1; \mu' ) < N (\sigma,  k^+-1; \mu (\sigma,  k^+-1))\Longrightarrow \nonumber \eeq
\beq \Longrightarrow  d^2 (\sigma,  k^+-1; \mu')= W (   N (\sigma,  k^+-1; \mu (\sigma, k^+-1)) - N (\sigma,  k^+-1; \mu' ) ) + \nonumber \eeq
\beq + \min_{ \{j_1, \cdots, j_{N (\sigma,  k^+-1, \mu (\sigma,  k^+-1))} \vert \forall i \leq N (\sigma,  k^+-1, \mu') \; , \; \vert 2 \nu ((\sigma,  k^+-1)+ \delta_{ k^+-1;j}) - \mu (\sigma,  k^+-1) - \mu' \vert < \vert \epsilon \}  } \eeq
\beq \sum_{j=1}^{N (\sigma,  k^+-1, \mu')} (\partial_{ k^+-1;j} B^{\mu} - \partial_{ k^+-1;j} B^{\mu})(\partial_{ k^+-1;j} B_{\mu} - \partial_{ k^+-1;j} B_{\mu})\nonumber \eeq
\beq K_{\mu} (\sigma,  k^+-1) = \nonumber \eeq
\beq = \bigg( \sum_{\{j \vert \forall i (K_j^{-2} (erfc \; (2 \nu ((\sigma,  k^+-1) + \delta_{k^+-1;j}) - \mu (\sigma,  k^+-1)))^{-1} \geq  erfc \; (2 \nu ((\sigma,  k^+-1) + \delta_{k^+-1;i}) - \mu (\sigma,  k^+-1)))^{-1} - \epsilon} \eeq
\beq e^{- \alpha d^2 (\sigma,  k^+-1; 2 \nu((\sigma,  k^+-1)+ \delta_{ k^+-1;j}) - \mu(\sigma,  k^+-1))}  \vert \psi ((\sigma,  k^+-1) + \delta_{ k^+-1;j}) \vert^2 \bigg)^{-1} \nonumber \eeq
\beq \alpha (\sigma,  k^+-1) \approx  K (\sigma,  k^+-1)  \exp (- d^2(\sigma,  k^+-1;  \nonumber \eeq
\beq (erfc^{-1} (M ((erfc \; (2 \nu ((\sigma,  k^+-1) + \delta_{k^+-1;1}) - \mu (\sigma,  k^+-1)))^{-1}, \cdots,  \eeq
\beq (erfc \; (2 \nu ((\sigma,  k^+-1) + \delta_{k^+-1;M (\sigma, k^+-1)}) - \mu (\sigma,  k^+-1)))^{-1}; \nonumber \eeq
\beq ; \; K^{-2}_{2 \nu ((\sigma,  k^+-1) + \delta_{k^+-1;1}) - \mu (\sigma,  k^+-1)} (\sigma, k^+-1)\, \cdots, \nonumber \eeq
\beq K^{-2}_{2 \nu ((\sigma,  k^+-1) + \delta_{k^+-1;M (\sigma, k^+-1)}) - \mu (\sigma,  k^+-1)} (\sigma, k^+-1)) ) e^{iS ( k^+-1)} \psi (\sigma, k^--1))^{-1} ) \big) \nonumber \eeq
\[ \frac{d \psi}{d \xi}= \left\{ \begin{array}{lll}
         \partial_{\sigma}^2 \psi + \partial_{\tau}^2 \psi & \mbox{if $(\sigma, \tau) \in D_{kl} \setminus \partial D_{kl}$};\\
        n^{\tau} \partial_{\tau} \psi  + n^{\sigma} \partial_{\sigma} \psi + \epsilon \alpha (\sigma, \tau) & \mbox{if $\tau = k^+-1$} \\
        n^{\tau} \partial_{\tau} \psi  + n^{\sigma} \partial_{\sigma} \psi - \epsilon \psi  (\sigma, \tau) & \mbox{if $(\sigma, \tau) \in \partial D_{kl}$ and $k^+ -1< \tau  \leq k^-$}.\end{array} \right.  \]
where $M (\sigma, k^+-1)$ is the total number of ''bridges'' that ''touch'' a point $(\sigma, k^+-1)$. While we still have ''d'' in the above equations, we have replaced $d (D_{kl}, D_{kj})$ with $d(\sigma, k^+-1; \nu)$. The role of "region" $D_{kl}$ has been replaced by a "point" $(\sigma, k^+-1) \in D_{kl}$, whereas the role of the region $D_{kj}$ has been replaced by its ''messenger'' $\nu$ that has ''diffused'' across the ''bridge'' which ''connects'' the ''earlier edge'' of $D_{kj}$ with an ''earlier edge'' of $D_{kl}$. Thus, $d(\sigma, k^+-1; \nu)$ is \emph{locally} defined at $(\sigma, k^+-1)$, as evidenced by the fact that $\mu$ is merely a real number. By inspection one finds that everything else above is likewise locally defined in $(\sigma, k^+-1)$. The dynamics of $\psi$ is depicted in the last equation where $\psi$ ''diffuses'' across the region $k^+-1 \leq \tau \leq k^-$ with ''source'' $\alpha$ located at $\tau = k^+-1$. The fact that the location of the source is $\tau = k^+-1$ as opposed to $\tau = k^-$ is the ultimate reason why $+ \tau$ ends up ''appearing as time'' as opposed to $- \tau$. One should not be intimidated by $-1$ since the ''local'' diffusion process provides ''spreads'' whatever is at $\tau=k^+-1$ to $\tau = k^-$

\subsection*{5. Conclusion}

In Section 2 we have shown how to identify probability amplitude of complex process in terms of wave function string surface without resorting to configuration space. This was done by claiming that the value of wave function is constant throughout each of the connected domains we have specified; thus, the value of wave function at any given point on the said domain represents the probability amplitude of the entire process pictured by that domain. Furthermore, we have defined the non-local dynamics of said wave function that produced the familiar probability amplitudes. Then in Section 3 we proceeded to extend this to GRW model; in other words we have ''translated'' the familiar GRW model in point particle case into the context of string theory, equipped with our new notion of probability amplitudes. Finally, in Section 4 we proposed a mechanism of ''converting'' the non-local laws we have postulated to local ones by means of invoking extra ''time'' coordinate $\xi$ so that, at $\xi \rightarrow \infty$ equilibrium, the desired ''non local'' laws arise. 

One feature of our theory is that we have not described a dynamics of shape of the string. The shape is said to be afore-given while the dynamics solely involves assigning probability amplitudes that each given ''part'' of that string ''takes place''. In other words, the ''afore-given'' shape of the string really represents ''all possible'' trajectories as opposed to one single trajectory; and then ''probability amplitude'' $\psi$ tell us which of these ''possible trajectories'' actually ''take place''  (although from ''philosophical'' point of view we believe that the said afore-given shape is ''reality'' while ''partial shapes'' is something we ''read off'' from ''field'' $\psi$). Naturally, the fact that the shape of the ''big string'' is still ''finite'' implies some form of discretization of the path integral. Thus, the fact that the shape of the string is afore-given is logically similar to ''afore-given'' structure of discrete spacetime in the discretized models of point particle QFT, at least in the gravity-free toy models (but, of course, in our case the spacetime is continuous, and the role of discretization is played by the finite number of string branches). 

Nevertheless, the fact that ''afore-given'' string stretches out throughout all values of $\tau$ and includes the behavior of $X^0$ implies lack of determinism in both $\tau$-time and $x^0$-time in a sense that we were not guided by any dynamical laws in postulating that structure. At the same time, the determinism in $\xi$-time is preserved: once we know the structure at $\xi=0$, we know that the structure will be the same for all other $\xi$, while the probability amplitudes evolve deterministically in $\xi$ as was shown in Section 4. In the discrete models of point particle QFT similar situation might arise. If we have a toy model without gravity, we might want to postulate ''afore-given'' lattice in Minkowski space. If that lattice is to stretch through both space and time, we are not guided by any dynamics of precise location of lattice points. Thus, we have an element of randomness in time, although it has negligible effect.

On a good side, the presence of more than one time coordinate allows us to "go back" in $x^0$ and $\tau$ as long as we go forward in $\xi$; this allows the  $x^{\mu}$-based relativity to co-exist with quantum mechanics, as stated by Nikolic' in \cite{Nikolic}. For example, this addresses grandfather's paradox. Of course, the relativity can not be respected with respect to a coordinate that we \emph{do} view as ''true time'' which in our case is $\xi$. Besides, we also violated relativity in $(\sigma, \tau)$ when we postulated $\tau= const$ ''lines'' to define our domains and also when we used plus signs in the ''diffusion equation'' in previous section. Nevertheless, we do preserve the relativity in $x^0$. That feature of "going back" in time in connection to string theory was also noticed by Nikolic'. He likewise utilized this feature to come up with interpretation of quantum mechanics on the string (see \cite{NikolicStrings}); although his interpretation is Bohmian while ours is GRW. Nevertheless, we are now leaving the question of why the dynamics is ''mostly'' in the future direction of $x^0$ unanswered; it might be a result of ''afore-given'' string structure that ''happens'' to statistically correlate $+ \tau$ with $+X^0$, for which we have no justification.

\end{document}